\documentclass[12pt,preprint,authoryear]{aastex}

\newcommand {\kms}{\mbox{km~s$^{-1}$}}
\newcommand {\msun}{\mbox{M$_\odot$}}

\slugcomment{Version  01/02/08 2:40pm}
\shorttitle{HRL Masers in MWC349A}
\shortauthors{Weintroub et al.}

\begin{document}
\title{SMA Imaging of the Maser Emission from the H30$\alpha$ Radio Recombination Line in MWC349A}

\author{Jonathan Weintroub, James M. Moran, David J. Wilner and Ken H. Young}

\affil{Harvard-Smithsonian Center for Astrophysics, 60 Garden Street, Cambridge, MA 02138}
\email{jweintroub@cfa.harvard.edu}

\author{Ramprasad Rao}
\affil{Institute of Astronomy and Astrophysics, Academia Sinica, 645 N. A'ohoku Place, Hilo, Hawaii 96720 }

\author{Hiroko Shinnaga}
\affil{Caltech Submillimeter Observatory, 111 Nowelo Street, Hilo, HI 96720}

\begin{abstract}
We used the Submillimeter Array to map the angular distribution of the
H30$\alpha$ recombination line (231.9 GHz) in the circumstellar region of the peculiar star MWC349A. The resolution was $1\farcs2$, but because of high signal-to-noise ratio we measured the positions of all maser components to accuracies better than $0\farcs01$, at a velocity resolution of $1~\kms$. The two strongest maser components (called high velocity components) at velocities near $-14$ and $32~\kms$ are separated by $0\farcs048 \pm 0\farcs001$ (60 AU) along a position angle of $102 \pm 1\arcdeg$.  The distribution of maser emission at velocities between and beyond these two strongest components were also provided. The continuum emission lies at the center of the maser distribution to within 10 mas. The masers appear to trace a nearly edge-on rotating disk structure, reminiscent of the water masers in Keplerian rotation in the nuclear accretion disk of the galaxy NGC4258.  However, the maser components in MWC349A do not follow a simple Keplerian kinematic prescription with $v \sim r^{-1/2}$, but have a larger power law index.  We explore the possibility that the high velocity masers trace spiral density or shock waves.  We also emphasize caution in the interpretation of relative centroid maser positions where the maser is not clearly resolved in position or velocity, and we present simulations that illustrate the range of applicability of the centroiding method.
\end{abstract}

\section{Introduction}

The hydrogen recombination line emission from the peculiar B[e] star MWC349A is double-peaked at frequencies above 100~GHz, and is thought to be due to maser action in a gaseous rotating circumstellar disk.  Thermal continuum emission from radio to IR wavelengths reveals that the star also has an ionized outflow.  Maser action occurs in recombination lines from about 100~GHz (H39$\alpha$) to 1800~GHz (H15$\alpha$) and beyond \citep{str96a, str96b}.  A physical model developed by \citet{thu94a,thu94b} suggests that the maser spots originate on the ionized surface of a rotating disk that is nearly edge-on in orientation.  \citet{pon94} performed detailed radiative transfer modeling and concluded that the masers are probably unsaturated with a maximum negative opacity (gain) of about five. 

High resolution VLA continuum images of MWC349A \citep[e.g.,][]{whi85, tafo04} show an hourglass-shaped image whose angular size scales with frequency $\nu$ as $\nu^{-0.7}$ and flux density as  $\nu^{0.6}$, as expected for a dense ionized outflow.  The star is located in the waist of an hourglass-shaped gas structure, with the north-south lobes corresponding to outflows, and the putative rotating disk presumed to  be located  around the star and viewed nearly edge-on.  It is interesting to note that VLA continuum images spanning the period of 1996--2006 \citep{rod07} show that the continuum morphology of the object has evolved from an hourglass structure to a more of a rectangular projected structure (i.e., the waist of the hour glass structure is no longer prominent). The spectrum of the maser emission is also variable \citep[see][]{mar89,gor01}.

\citet{pla92} observed the H$30\alpha$ maser emission and showed that its dominant characteristic is two features (aka ``spots'') at velocities of $-14$ and $32~\kms$ separated by about $0\farcs065 \pm 0\farcs005$, along a position angle of about $107\arcdeg \pm 7\arcdeg$.  The close correspondence between the position angle of the spots and that of the waist of the hourglass continuum map suggests that the maser spots are located in the waist of the continuum.  For a distance of 1200 pc \citep{coh85}, the velocity and spatial separation between the two maser peaks of $47~\kms$ and 120 AU respectively, and a central mass of about $30~\msun$, can be inferred from Kepler's third law  \citep[e.g.,][]{str96a}. 

\begin{figure}
\begin{center} 
\includegraphics[angle=-90,scale=0.8]{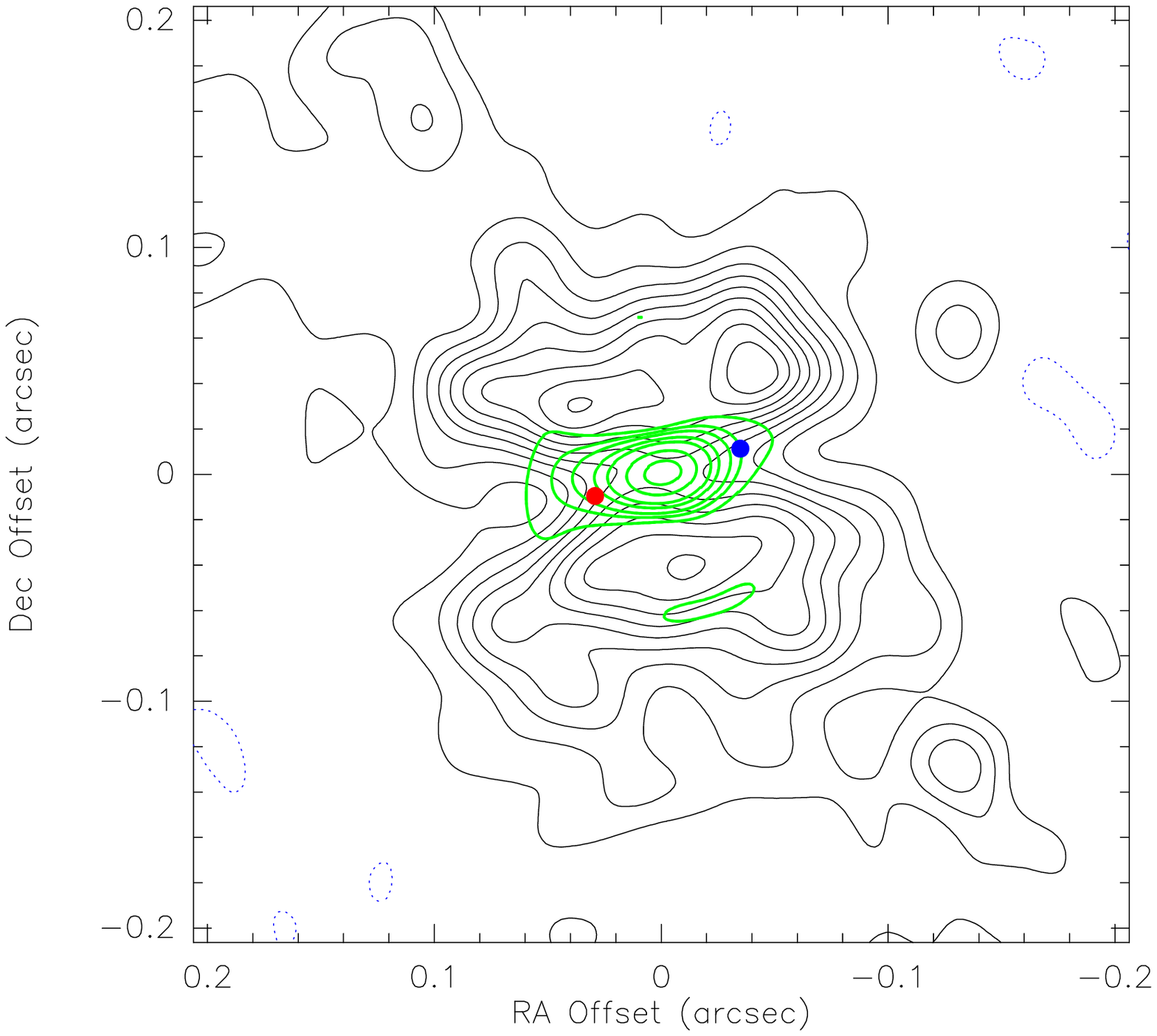}
\caption{Superposition of the 7--mm VLA continuum image (black contours) made by \citet{tafo04}, $2.27~\micron$ Keck aperture-masked image (green contours) made by \citet{dan01}, and two maser components (red dot for 32~$\kms$ and blue dot for $-16~\kms$) measured by \citet{pla92}. Relative astrometric accuracy among the wavelength bands was poorer than 0$\farcs$1, we have co-registered the images to maximize the probable symmetry of the source.  This figure is an updated version of a similar figure published by \citet{dan01}. The reference position for the image is RA$ = 20^h 32^m 45.528^s$ and DEC$ = 40\arcdeg 39\arcmin 36\farcs622$ (2000), taken from the VLA observations.
\label{danchicont}}
\end{center}
\end{figure}

Infrared continuum images at 1.65, 2.27, and $3.08~\micron$ were made by \citet{dan01} with the Keck telescope using an aperture masking technique.  The images are elliptical with major axes of 36, 47 and 62 mas, respectively.  
The position angles of the images are about $100\arcdeg$.  These images trace the dust and the neutral gas that is probably undergoing photoevaporation, giving rise to the whole hourglass ionized region \citep{hol94}. The structures of the radio continuum at 7~mm \citep{tafo04}, the infrared at $2.27~\micron$, and the maser emission from OVRO \citep{pla92} are shown in a composite image in Figure \ref{danchicont}.

%FIGURE 1

Here we present interferometric images of the maser emission in the H$30\alpha$ radio recombination line, which have much higher signal-to-noise ratios than those of previous investigations.  This allows for the study of the distribution of maser emission in significantly greater detail.  We have additional data from the SMA in the H$26\alpha$ (353 GHz) and H$21\alpha$ (652 GHz) transitions.  These results have lower signal-to-noise ratio than the ones presented here. They will be discussed in a later publication.

\section{Observations and Data Reduction}

The H30$\alpha$ observations made on 2004 Sept 12 are presented in this paper.  The weather was excellent for 230~GHz operations with $\tau_{225}$ varying between 0.05 and 0.07 during the night. The array was in the ``extended'' configuration with seven antennas in operation (pads 1, 8, 11, 14, 15, 16 and 17) providing baselines from 21 to 156 m. The synthesized beam at 230 GHz was 1$\farcs$2 by $0\farcs$9 with a PA of 90$\arcdeg$.  The correlator was configured in a hybrid mode.  One 82 MHz ``chunk'' was centered on the maser emission and fed to a 512-channel cross correlator. The other 23 chunks were set at frequencies surrounding the maser band and fed to a bank of 32 channel correlators. The spectral resolution on the maser chunk was formally 0.24 MHz, but the data were averaged into bins at intervals of 0.76 MHz, exactly $1~\kms$, covering a velocity range of $-53$ to $53~\kms$. 

The observing schedule interspersed observations of MWC349A with the quasar calibrators 2013+370, 2202+422 and 2232+117. The SMA correlator was configured to take 30 second scans.  Our observing cycle consisted of 24 scans on MWC349A, followed by 6 scans on each of the three quasar calibrators. 2232+117 became available a little later than the other two quasars, and was the only quasar calibrator in the sky at the end of the track. After MWC349A set, we took 50 minutes of data on Titan for absolute flux calibration. The total integration time on MWC349A was 4.8 hours.

Bandpass calibration was a crucial aspect of these observations, since the angular distribution of the maser emission is only about 0$\farcs$05, or one-twentieth of the synthesized beam.  Hence, after MWC349A set, over three hours of additional bandpass calibration data on Venus and Uranus was taken.  We tested the stability of our bandpasses by dividing the Venus data sets into two parts and calibrating one half with the other. We then did a vector average of the visibility spectra on all baselines in the calibrated portion of the data. The phase deviation from zero was less than $\pm 0.5\arcdeg$ across the band. This means that any systematic error in relative positions is less than 1/720 of the resolution, or about $\pm0\farcs001$.  

Data reduction was carried out with the software packages MIR (for flagging, bandpass, gain, and flux calibration) and Miriad (for inverting the visibilities to form line images and applying the CLEAN algorithm). In MIR, data were first weighted by system temperature to account for sensitivity variations due to elevation and atmospheric changes.  The bandpass calibration used all three quasars and three hours of high signal-to-noise ratio (SNR) data from Venus.  The gain calibration used the interspersed data from the quasar 2013+370.  Absolute flux calibration was achieved on the data from Titan, whose flux density was estimated with the SMA's online visibility calculator as 1.2~Jy.  The MWC349A data were exported in FITS file format for further manipulation in Miriad.  The latter program was used to invert the visibilities to create line and continuum images.  The Miriad implementation of the CLEAN algorithm was used to deconvolve the dirty beam from the images. The vector-averaged visibility spectrum is shown in Figure \ref{spectrum}. The rms accuracy of our images at $1~\kms$ resolution was about 125 mJy. 

\section{Astrometry and Continuum Measurement}

When making these observations, the phase center of the array was set to the published coordinates of MWC349A from the Simbad database: RA$ = 20^h 32^m 45.53^s$ and DEC$ = 40\arcdeg 39\arcmin 37\farcs0$ (2000). These coordinates came from the catalog of H-alpha emission stars in the Northern Milky Way published by \citet{kaw99}.  The coordinates we derive for the centroid of the 1.3 mm continuum emission are RA $= 20^h 32^m 45.53^s \pm 0.01$ and DEC$ = 40\arcdeg 39\arcmin 36\farcs6 \pm 0.1$ (2000). There is significant discrepancy of 0$\farcs$4 in declination. Our position agrees well with the centroid position of the 7 mm continuum emission determined by \citet{rod07} of  
RA $= 20^h 32^m 45.528^s \pm 0.05$ and DEC$=40\arcdeg39\arcmin36\farcs622 \pm 0.005$ (2000) for epoch 2004.9. Since the positions of the radio images are coincident to within their errors $0\farcs1$, we suspect that the apparent offset of the H$\alpha$ position in declination is due to uncertainty in the optical measurement. All images in this paper are referenced to the 230 GHz position of the continuum emission. 

The spectrum of MWC349A closely follows the form of $\nu^{0.6}$, and the angular size proportional to $\nu^{-0.7}$, indicative of a square law decrease in density with radius. The flux density and size extrapolated from 43 GHz to 230 GHz are 2.1 Jy and 0$\farcs$04, respectively. The flux density we measured was about 2 Jy. The source was unresolved, as expected, by the SMA. The relative alignment of the continuum and maser emission is accurate to 0$\farcs$01. 

\section{Results}

%FIGURE 2
\begin{figure}
\begin{center} 
\includegraphics[angle=0,scale=0.6]{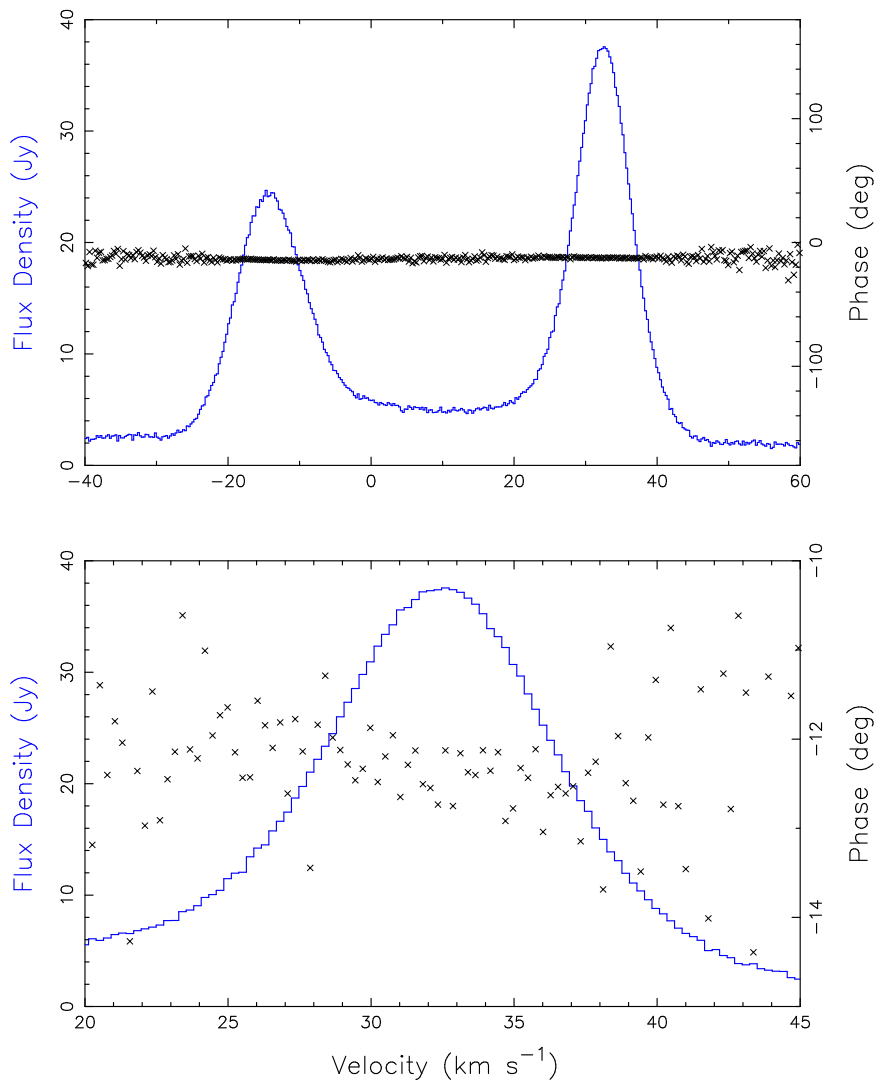}
\caption{(top) Vector averaged visibility spectrum  of all of the data on all baselines from the 2004 September 12 observations of the H30$\alpha$ line: bandpass, gain, and flux calibrations have been applied. (bottom) Detailed view of the visibility spectrum in the redshifted part of the spectrum. The scatter in the phase across the red feature is small because of the high SNR and angular proximity of the spectral components.  In both panels, crosses show the phase, and the continuous lines show the amplitude. The velocity axis refers to the local standard of rest and is based on a rest frequency of 231.9009~GHz. In this figure only the data are displayed at the intrinsic correlator resolution of $0.24~\kms$.  \label{spectrum}}
\end{center}
\end{figure}

We assumed that the image at each velocity could be modeled as a point source. To determine the centroid position, a two-dimensional Gaussian function was fitted to each image.  A discussion of systematic errors that can result from this technique due to multiple unresolved features is described in Section 5.4. Briefly, we found that if the spectral widths of individual masers are appropriate for unsaturated amplification with a gain of about 100 or more, then the position centroiding analysis (PCA) method will give unbiased results.  If, however, the high velocity emission near $-14$ and $32~\kms$ is dominated by single components with amplification gain factors of only about five, then the distribution of components around the velocities will be biased and the inferred power law indices will be larger than the Keplerian value of 0.5.

The formal error in the position estimate derived by the PCA method is given by the equation 
\begin{equation}
\Delta\theta = \frac{1}{2} \frac{\theta_r}{SNR}~~,
\end{equation}

\noindent 
where $\theta_r$ is the angular resolution (beamwidth) of the array and $SNR$ is the signal-to-noise ratio in the imaged velocity channel. Hence, the accuracy of PCA method can be much greater than the beamwidth for high values of $SNR$.  For the strongest feature of 42 Jy, with rms noise of 125 mJy and resolution of 1$\farcs$2, the formal centroiding accuracy is $0\farcs0018$. The spectrum in Figure \ref{spectrum} (bottom) shows that the rms phase noise is about $1\arcdeg$. This is about an order of magnitude better than the accuracy achieved by \citet{pla92}.

%FIGURE 3
\begin{figure}
\begin{center}
\includegraphics[angle=0,scale=0.7]{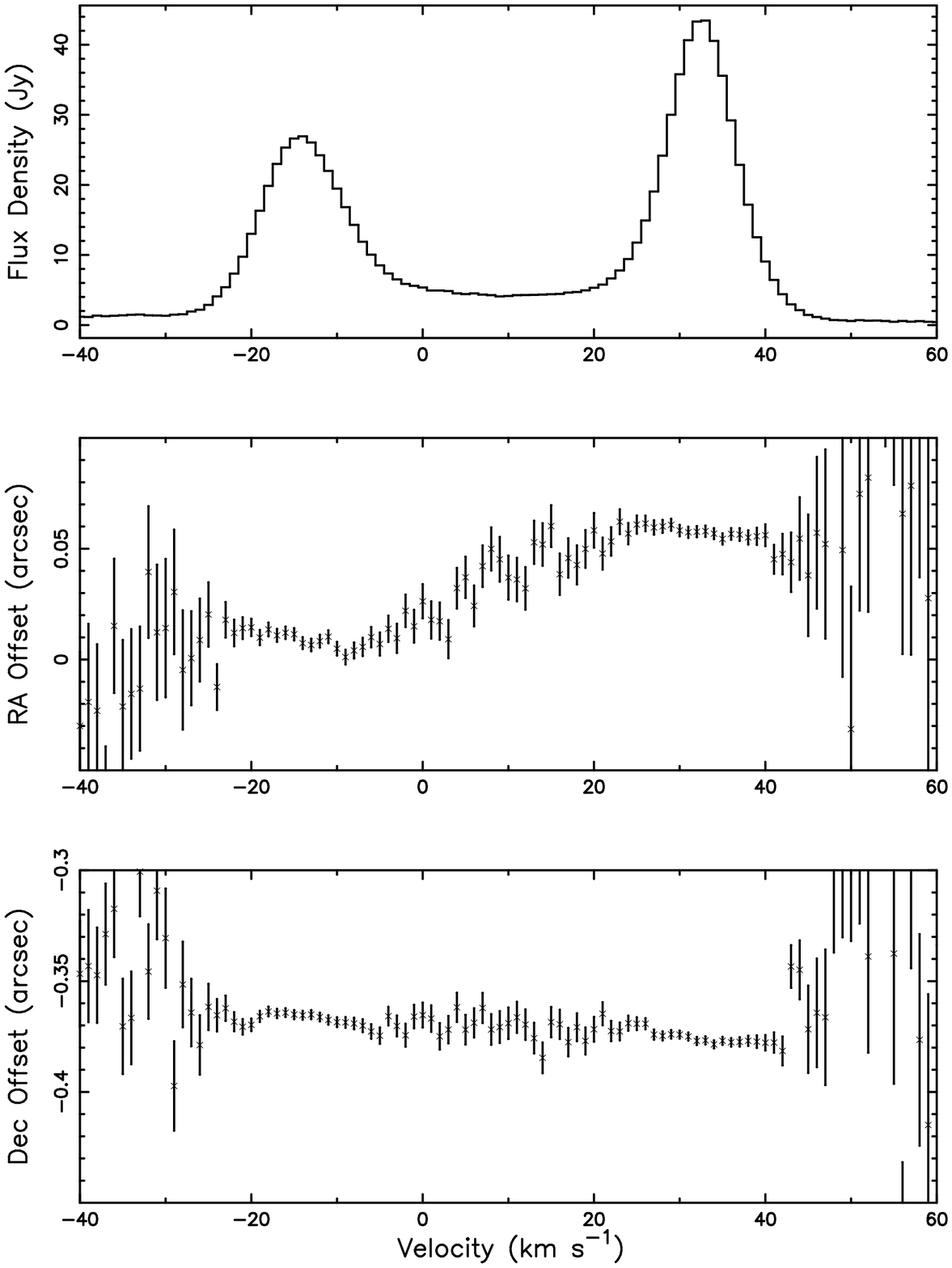}
\caption{Results of the centroiding analysis applied to each velocity channel. (top) Spectrum, i.e., the peak amplitude at each velocity. (middle and bottom) RA and declination positions at each velocity with respect to the phase center of the Array. The continuum source is at relative position: $\Delta$RA = 0$\farcs031$ and $\Delta$DEC = $-$0$\farcs369$.  The position errors shown are based on the PCA method, and closely approximated by equation (1).
\label{spotvelh30}}
\end{center}
\end{figure}

Figure \ref{spotvelh30} shows the spectrum and the position associated with each velocity channel from the centroiding analysis. Notice how the RA trends downwards in the blueshifted component, reverses and trends upwards in the trough between the components,  then  reverses and trends downwards again through the redshifted peak.  No such reversals are apparent in the corresponding declination plot.  

%FIGURE 4
\begin{figure}
\begin{center} 
\includegraphics[angle=-90,scale=0.4]{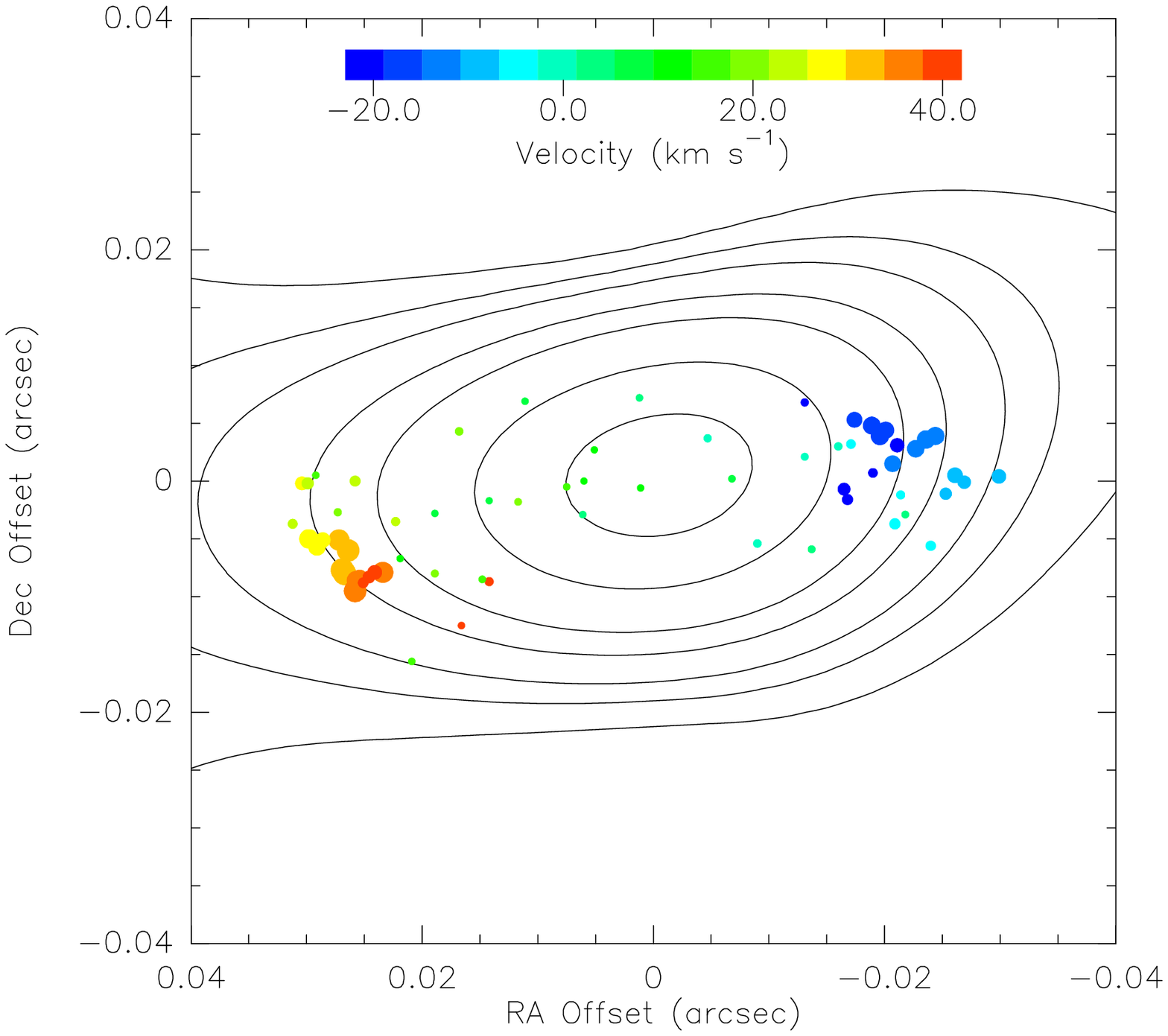}\\
\vspace{2mm}
\includegraphics[angle=-90,scale=0.4]{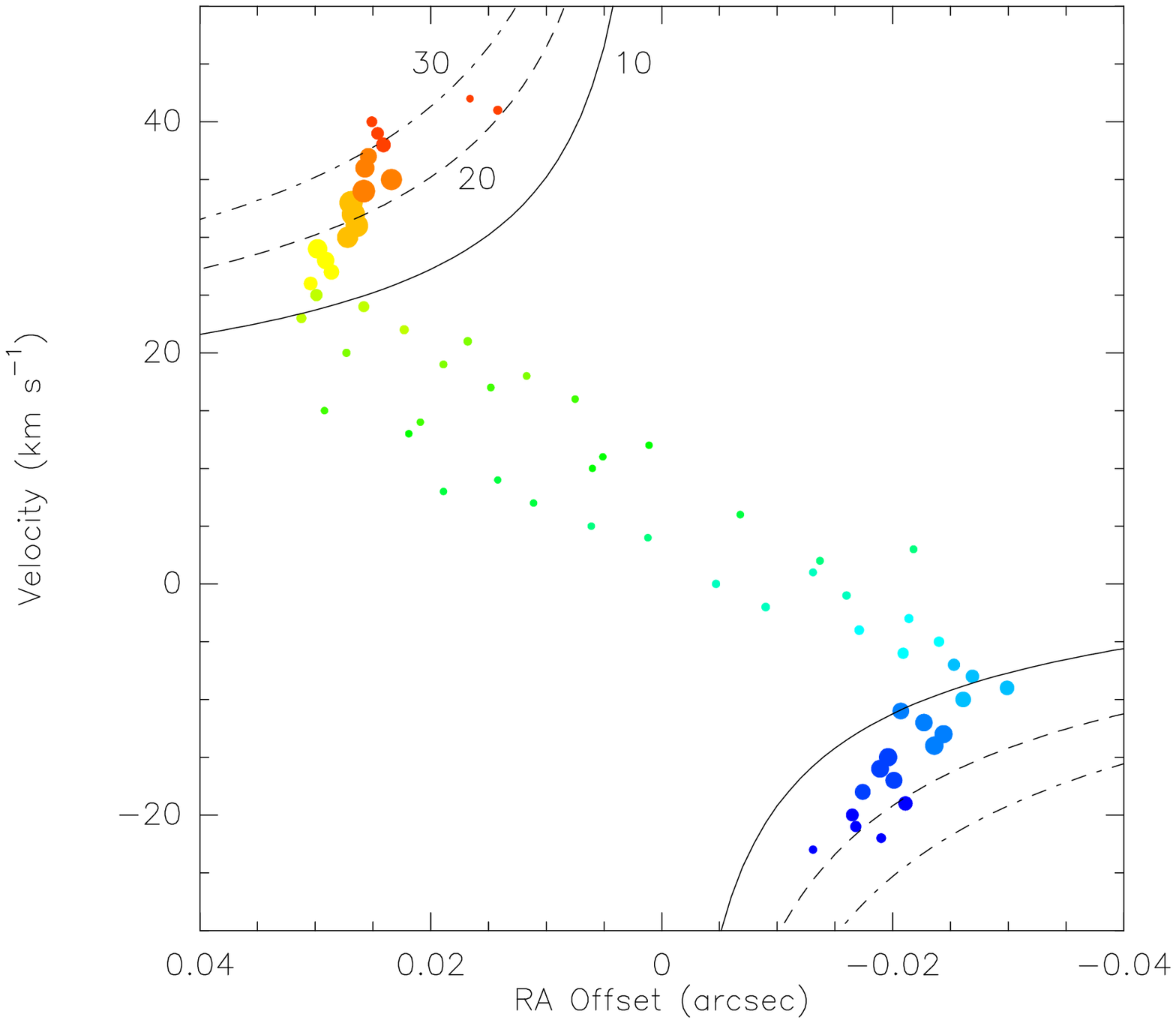}
\caption{(top) Distribution of maser spots with respect to the position of the unresolved 230 GHz continuum emission. This relative alignment is accurate to $0\farcs01$. An angular offset of $0\farcs05$ corresponds to a linear dimension of 60 AU or $0.9 \times 10^{15}$ cm for a distance of 1200 pc. Areas of the spots are proportional to their flux densities and color coded by velocity.  Error bars, shown in Figure \ref{spotvelh30}, are omitted here for clarity. Contours represent the 2.27 $\micron$ infrared emission from \citet{dan01}. The IR image is nominally aligned with the continuum position. (bottom) The p-v diagram, where right ascension is used as the position axis. High velocity data points (velocities in the range of 21 to $42~\kms$ and $-5$ to $-21~\kms$) are overlaid with Keplerian curves for central masses of 10, 20 and $30\msun$. The high velocity wings are considerably steeper than expected for Keplerian motion at a fixed azimuth angle.  
\label{spotvrah30}
\label{3masskepler}}
\end{center}
\end{figure}

Figure \ref{spotvrah30} shows the distribution of masers as a function of velocity (aka a ``spot'' map). Note that every velocity channel with significant flux density is shown.  No attempts have been made to identify specific spectral features. Figure \ref{spotvrah30} also shows a position-velocity (p-v) 
plot of velocity vs. right ascension, approximately the major axis of the distribution. In the discussion that follows, we refer to the emission in the spectral peaks (i.e., $<0$ and $>20~\kms$) as ``high velocity'' emission, and the spectral range in between where the position changes linearly with 
position, as the ``low velocity" emission. 

We can compare our results quantitatively with those of \citet{pla92}. They measured the mean separation between emission in the range $32 \pm 10~\kms$ and that in the range $-16 \pm 10~\kms$ to be $0\farcs065 \pm 0\farcs005$ at a position angle of $107\arcdeg \pm 7\arcdeg$.  If we average our measurements over the same range (amplitude weighted) we obtain a separation of $0\farcs048 \pm 0\farcs001$ at a position angle of $102 \pm 1\arcdeg$. 

\section{Discussion}

\label{disc}

%FIGURE 5
\begin{figure}
\begin{center}
\includegraphics[angle=0,scale=0.7]{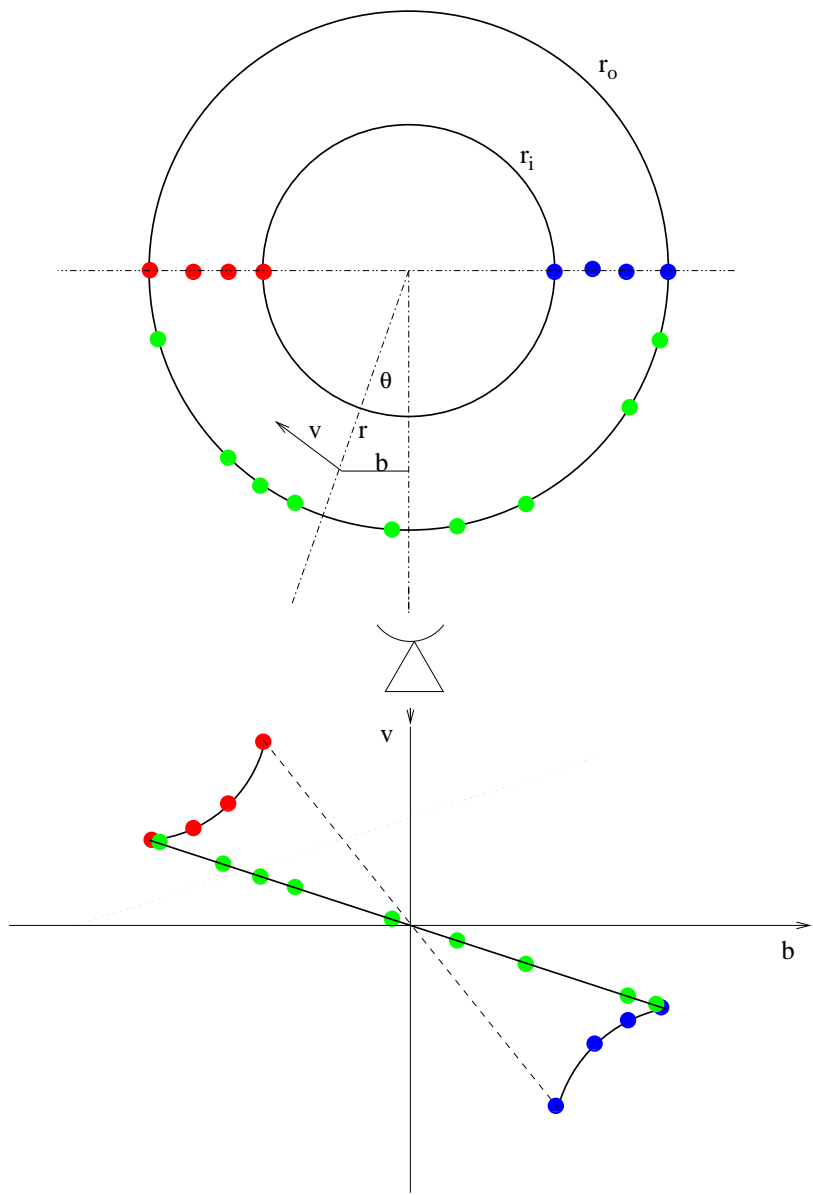}
\caption{Sketch showing the maser positions within an edge-on unfilled disk in Keplerian rotation.  The small filled circles represent the masers. Their positions are shown within the disk, and corresponding positions on the position-velocity diagram.  Were the masers randomly distributed, they would randomly fill the characteristic ``bowtie'' envelope; however, the shape of the measured position-velocity diagram of Figure \ref{spotvrah30} suggests that the masers are concentrated on the outer edge of the disk, with the curved Keplerian segments corresponding to the masers along the midline (the line passing through the star at the center of the torus, and perpendicular to the line of sight). 
\label{cartoon}}
\end{center}
\end{figure}

We discuss possible dynamical models to explain the position-velocity data that we have obtained. We begin by reviewing the characteristics of a Keplerian disk to frame the discussion. We find that the high velocity features follow a power-law distribution of the form $v \sim r^{-\alpha}$, where $\alpha$ is much larger than the Keplerian index of 0.5. We then show that the large power-law index can be understood if the masers are in Keplerian motion, but distributed along spiral arms.  Finally, we present simulations to show the reliability of the centroiding method of determining relative maser positions to small fractions of the spatial resolution. 

\subsection{A Thin Annular Keplerian Disk}

Figure \ref{cartoon} shows the characteristic p-v relationship of a set of discrete masers in purely Keplerian rotation.  For an edge-on disk full of compact masing sources in Keplerian rotation, a maser spot at radius $r$ has a velocity $v$ and a line-of-sight velocity $v_z$ of (see Figure \ref{cartoon}):

\begin{equation}
v_z=v \sin{\theta} = \sqrt{{GM}\over{r}} \sin{\theta}~~,
\end{equation}

\noindent
where G is the gravitational constant, $M$ is the stellar mass, and $b$ is the projected offset from the star or impact parameter. Since $\sin\theta = b/r$, $v_z$ will be proportional to $b$ for masers at fixed values of $r$, i.e., 

\begin{equation}
v_z = \sqrt{{GM}\over r^3} ~b~~.
\end{equation}

\noindent
In practical units equation 3 can be written as $v_z = 27(M/r^3)^{1/2} b$ for a distance of 1200 kpc when $v_z$ is in $\kms$, $b$ and $r$ are in mas, and $M$ is in $\msun$.  Masers at a fixed radius will extend along a straight line between the Keplerian limits where $b = \pm r$. For a disk with maser emission bounded by an inner and outer radii of $r_i$ and $r_o$, the masers will lie within a ``bowtie'' region in the position-velocity diagram, bounded by the straight lines defined by $v_z=\sqrt{GM/r_i^3~}b$ and $v_z=\sqrt{GM/r_o^3}~b$, respectively (see Figure \ref{cartoon}).  Features lying along the curved outer boundaries arise on the ``midline'' where $\theta  = \pm 90\arcdeg$.

The distribution of masers in the actual p-v diagram suggests the masers are in three regions: approximately along the midline, one region on either side of the star, and in a relatively thin annular region of radius equal to about the outer radius (0$\farcs$025, or about 30 AU) as traced by the high velocity masers.  It is not possible to tell whether masers in this annular region come from the front side or back side of the disk.  However, if they are spread out on both sides, then the ratio of the minor to the major axis, which is about 0.25, suggests that the disk is tilted by about 15$\arcdeg$ to the line of sight. If the front side of the disk is tipped down, as suggested by \citet{rod94} with respect to us, there appears to be a deficit of masers on the front 
side of the disk (see Figure \ref{spotvrah30}).  The lower panel of Figure \ref{3masskepler} shows the p-v plot with Keplerian curves for masses of 10, 20 and 30 M$_{\odot}$.  The actual distribution of masers in the diagram crosses these three curves; hence, it is difficult to assign a specific mass without a more complex model (see Section 5.4). 

\subsection{A Power Law Velocity Curve}

We can characterize the ``Keplerian'' wings by a power law model of the form 

\begin{equation}
v_z(b)=v_0+ k(b-b_0)^ {-\alpha}~~.
\end{equation}

\noindent
We have used the RA positions for $b$. Since this equation is nonlinear in the parameters, we solved for their values through a chi-square minimization technique. The errors were estimated from the offset from the global minimum necessary for $\chi^2$ to increase by 1. The parameters and their errors are:

$v_0 = 2 \pm \kms$;

$b_0 = 0\farcs005 \pm 0\farcs002$;

$k = 0.00014 \pm 0.00003$; and

$\alpha = 3.2 \pm 0.1$.

\begin{figure}
\begin{center} 
\includegraphics[angle=-90,scale=0.5]{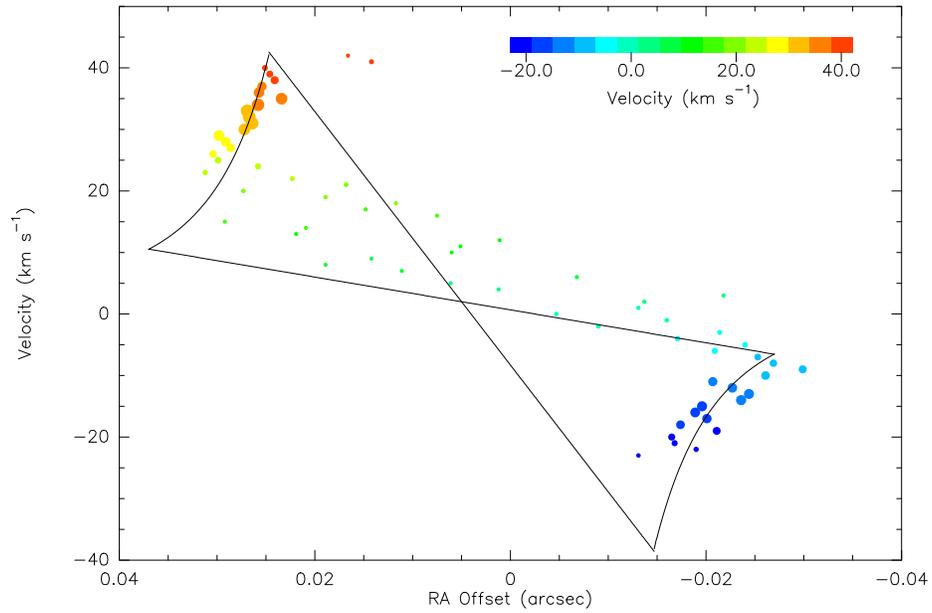}
\caption{Power law model fit to the high velocity portions of the position-velocity data. Straight lines of the bowtie have been set to match the limits of the high velocity emission and correspond to radii of 0$\farcs$019 and 0$\farcs$032, respectively. The fact that some features lie outside the bowtie area could be an indication of the thickness of the disk or its inclination.
\label{hvfit}}
\end{center}
\end{figure}

\noindent
The fitted power-law rotation curve is shown on the p-v data in Figure \ref{hvfit}.  A bow-tie diagram has been superimposed whose linear limits have been set according to the limits of the high-velocity features. Note that the blueshifted features have more curvature in the p-v diagram than the redshifted features.  Because of the symmetry imposed in the fit (i.e., both sides of the p-v diagram) are assumed to have the same power law index, the centroid of the p-v diagram is moved to the apparently off-centered location as shown.  The systemic velocity is $2~\kms$ in this model, significantly offset from the mean velocity of the maser emission of $9~\kms$ and the central position is 
0$\farcs005$ offset from the continuum peak. The masers lie in the angular region of radii of about $0\farcs019$ and $0\farcs032$.  Note, however, that the outer edge is defined by the blueshifted features while the inner edge 
is defined by the redshifted features, and neither side of the disk is completely filled.

This model seems somewhat unrealistic due to the irregular boundaries of the disk and unusual systemic velocity.  Positional symmetry could be restored if the power law indices for the red and blue features were allowed to be different. The power law model is unusual since the power law index is so large.   A possible explanation could be that the disk is undergoing a magnetically powered spin up. \citet{thu99} measured the line-of-sight magnetic field strength in the masing gas to be 22 mG.  If the masers lie on the midline of a rotating disk, this field would correspond to a toroidal component. Even if the mass of the disk is assumed to be 30~$\msun$, the kinetic energy is considerably less than both magnetic and thermal energy; hence, the magnetic field may control the dynamics. 

%FIGURE 6

\subsection{A Spiral Arm Model}

The velocities of the high velocity maser components may be consistent with Keplerian motion if they do not lie on a line of constant azimuth angle such as $\theta = \pm 90\arcdeg$ (i.e., midline). We consider the case where the maser velocities are Keplerian, but the values of $v_z$ deviate substantially because of the distribution of azimuth angles.  We assumed that the disk is edge-on and the central stellar mass is 30 M$_{\odot}$. Note that the p-v diagram for a 30 M$_{\odot}$ star falls outside the p-v data and just touches it at extreme velocities (see Figure \ref{spotvrah30}).  The azimuth angle of any maser with position $b$ and line of sight velocity $v_z$ from equation 2 will be given by

\begin{equation}
     \theta=\pm \sin^{-1}\left[\frac{b v_z^2}{GM}\right]^{1/3}~. 
 \end{equation}

\begin{figure}
\begin{center} 
\includegraphics[angle=-90.,scale=.5]{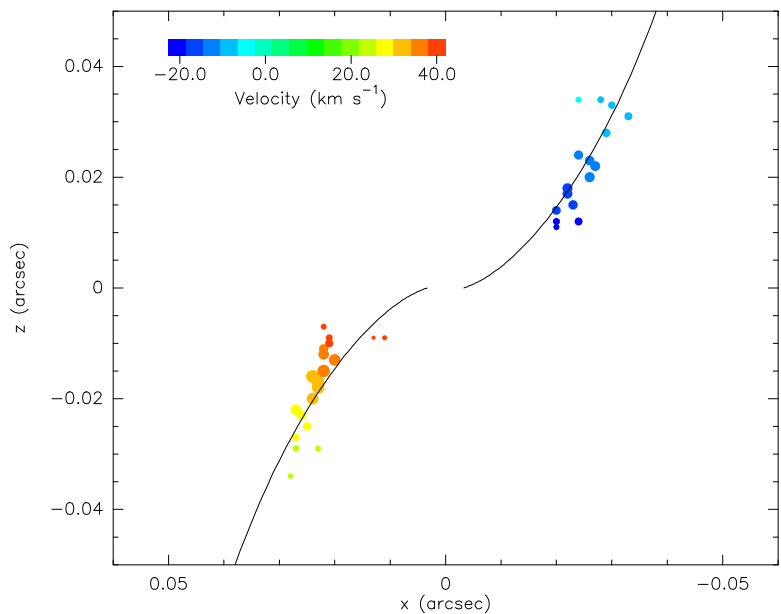}
\caption{``Top'' view of the maser distribution, i.e., seen from an axis perpendicular to the plane of the maser distribution. The direction to the earth is along the negative z-axis. The deprojection was accomplished by assuming that masers are in Keplerian orbits with the use of equation 5. The redshifted features and the blueshifted features were assumed to arise from in front of 
and behind the midline, respectively. The rapid change in azimuth angle with velocity can explain the steepness of the p-v plot. A possible model with trailing arms is plotted on the data.  This model requires that the PA of the disk be about $115^\circ$ (see Figure 4) and that the inclination be about $85^\circ$ in order to explain the declination distributions of the masers.}
\label{spiral}
\end{center}
\end{figure}

\noindent The position with respect to the midline, $z$, is then given by $z = b/\tan\theta$. Note that there is a two-fold degeneracy in this calculation, i.e., we cannot determine whether a particular maser is on the front side or back side of the midline from its velocity alone. We have chosen to put the red features on the near side of the disk, in order to allow a trailing spiral arm configuration. Note that a careful examination of Figure 4 shows there is a slight, but significant, trend for both the red and blue features to increase in declination as their velocities decrease. In order for this to be consistent with the spiral sense, the disk must be tipped up in front (i.e., $i=85 \arcdeg$) and have a PA of about $115 \arcdeg$. Note that the apparent PA of the masers is about $102 \arcdeg$. The large scale ionization structure suggests that the inclination angle is $95 \arcdeg$ (disk tipped down in front) \citep*{rod94}, and the position angle of the dust is about $100 \arcdeg$. These orientations would mean that the maser disk is misaligned with the dust structure and the large scale structure of the ionized gas by about $10 \arcdeg$ in both PA and $i$. An alternative model would be that the masers are tracing a leading spiral structure. The PA would still need to be about $115 \arcdeg$, but the inclination angle would be $95 \arcdeg$ (tipped down in front). We note that the spiral arms in most spiral galaxies are trailing. However, there are a few known examples of leading spiral arms \citep[e.g., NGC4622,][]{but03}. We favor the trailing spiral model with the front side of the disk tipped up. We selected a spiral structure of the form $r = r_0 e^{k(\theta-\pi/2)}$ to approximate the maser distribution.  The fitted spiral model is shown in Figure \ref{spiral}. 

%FIGURE 7

We will not explore or attempt to justify the existence of a spiral 
structure in the envelope of MWC349A. However, we mention that \citet{qui05} 
have developed a theory of spiral arms to explain the appearance of the 
circumstellar dust disks of HD100546 and HD141569. Note, however, that the 
Toomre Q parameter associated with the ionized disk of MWC349A is much greater 
than unity because the surface density of the ionized portion of the disk is 
so small. Hence, self gravity is not a likely explanation for spiral structure. That is, $Q = v_s \Omega/(G\Sigma$) where $v_s$ is the sound speed ($\sim10~\kms$), $\Omega$ is the angular rotation rate (for $r = 4 \times 10^{14}$ cm and $v = 32~\kms~\Omega = 5 \times 10^{-9}~s^{-1}$) and $\Sigma$ is the surface density (for $n_e = 10^7$ cm$^{-3}$ and thickness of $10^{14}$ cm, $\Sigma = 2.5 \times 10^{-3}$ g cm$^{-2}$). However, it may be possible that the stellar companion MWC349B, which is offset by 2$\farcs4$ at a position angle of $100 \pm 2\arcdeg$ \citep{coh85} (approximately at the position angle of the maser disk), could tidally disrupt the disk. 

\subsection{Evaluation of the Applicability of the Position Centroid Analysis (PCM) Method}

\label{simulation}

The technique of centroiding images to obtain positional measurements with accuracies much smaller than the instrumental resolution is well established in all wavelength regimes in astronomy. For an isolated source, with no confounding systematic measurement problems, the position can be found to the accuracy imposed by the instrumental noise limit (i.e., equation 1). This technique has been applied with particular effect in determining distributions of maser sources where the masing components are clearly distinguishable in both velocity and position \citep[see, for example][]{rei80}.  In this case each maser feature, which occupies a unique velocity range, appears as an unresolved ``spot.''  Even in the case where the features are well resolved in velocity only, but not in position, this technique is very effective.  For example, in NGC4258, the features are well separated in velocity over a range of $2000~\kms$ with line widths of about $1~\kms$, but not angular extent where the angular extent for the low velocity masers is 1 mas and the instrumental resolution is about 200 $\mu$as \citep*[see][]{mor99}.

In the case of MWC349A, the maser spectrum has a rather simple appearance, dominated by two almost perfectly Gaussian shaped profiles centered at 32 and $-14~\kms$. Although the instrumental resolution of the SMA is much finer than the linewidths of these features, their profiles do not reveal any substructure. Hence, the maser is resolved neither in velocity nor position, and the
interpretation of centroided images must be made with  caution and insight. Systematic artifacts, such as the appearance of a velocity gradient in a source that is actually composed of two components closely spaced in velocity and position is well known. For example, \citet{che06} have shown that a source of molecular emission (non-maser) in W3(H$_2$O), thought to be a gradient indicative of rotational motion, was in fact a double source whose components have slightly different velocities.

In this section, we report the results of the simulations we have made
for MWC349A following the methodology of \citet{che06} to investigate the effects of source blending. Since the structure we have discussed in this paper is essentially one dimensional (i.e., along a line with position angle of about 102$\arcdeg$), we have done only a simulation in one positional coordinate. We generated data for an image ``square'', i.e., intensity vs. velocity and position coordinate (x) from a set of Gaussian components. We then fit the image at each velocity with a Gaussian function and estimate its central position.  Note that the width of the Gaussian function in position is always very close to the instrumental beamwidth (1$\farcs$2 in our case), since the total maser extent is only 0$\farcs$06.

\begin{table}[bp]
  \caption{Parameters of 3 and 9 Component Simulations.\label{centroidstab}}
\centering   
\begin{tabular}{cccc}
\tableline\tableline
Three component model\tablenotemark{a}\\
\tableline
Amplitude  & Velocity & Width\tablenotemark{d} & Position \\
   (Jy)     & (km s$^{-1}$) & (km s$^{-1}$) & ($^{\prime\prime}$)\\
\tableline
23.7 & -14.1 & 4.7 & -0.025 \\
4.5 & 7.7 & 21.0 & 0 \\
40.2 & 32.4 & 4.0 & 0.025 \\
\tableline\tableline
Nine component model\tablenotemark{a}\\
\tableline
Amplitude  & Velocity & Width & Position \\
  (Jy)      & (km s$^{-1}$) & (km s$^{-1}$) & ($^{\prime\prime}$)\\
\tableline
3.5\tablenotemark{c} & -21.6 & 4.5 & -0.0117 \\
23\tablenotemark{c} & -14.7 & 4.5 & -0.0210 \\
9\tablenotemark{c} & -10.6 & 4.7 & -0.0319 \\
3.5\tablenotemark{b} & 0 & 5 & -0.0135 \\
3.5\tablenotemark{b} & 9 & 5 &  0.0025 \\
3.5\tablenotemark{b} & 18 & 5 & 0.0175 \\
9\tablenotemark{c} & 29 & 5.0 & 0.0355\\
42\tablenotemark{c} & 32.7 & 4.0 & 0.0260 \\
4\tablenotemark{c} & 40 & 4.0 & 0.0162 \\
\tableline
\end{tabular}
\tablenotetext{a} {The position-velocity diagrams for these models are shown in Figure \ref{centroids}.}
\tablenotetext{b} {These three components create the central ``gradient'' in the position-velocity diagram.}
\tablenotetext{c} {These components create the ``Keplerian'' wings in the position-velocity diagram. Their positions are appropriate for midline location, a central mass of 20 $\msun$, a systemic velocity of $9~\kms$ and a central position of $0\farcs0025$.}
\tablenotetext{d} {Widths ($\sigma_v$) are in natural units (FWHM = 2.3 $\sigma_v$).  For a thermal line, $\sigma_v = 0.091\sqrt{T_e}$ where $T_e$ is electron temperature.  Hence for $T_e = 10,000 K$, the thermal line width would be $9.1~\kms$, considerably broader than the model values of 4-5 $\kms$.  If, however, as suggested by \citet{pon94}, the masers are unsaturated with gain factures of about 5, the lines will appear narrowed by a factor of $\sqrt{5}$ , giving line widths of about $4~\kms$.} 
\end{table}

\begin{figure}
\begin{center} 
\includegraphics[angle=0,scale=0.6]{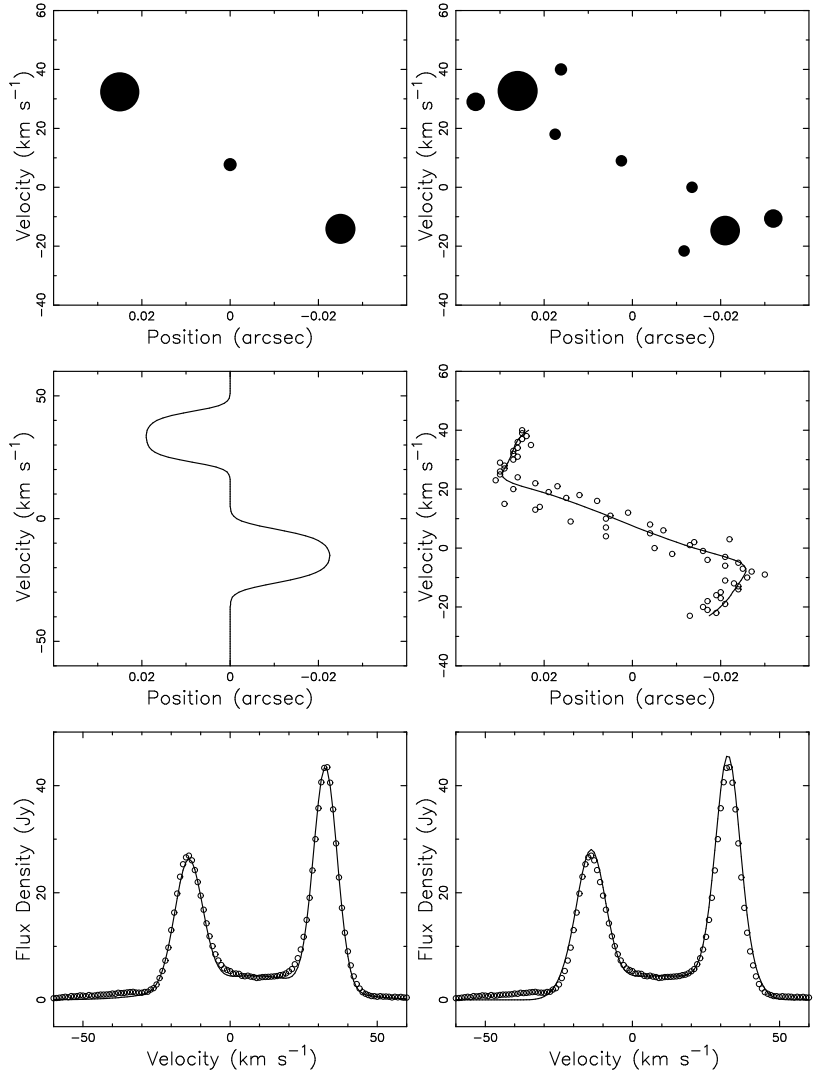}
\caption{Two simulations of the maser distributions in MWC349A from a model with components separated in only one spatial coordinate: (left) three-spot model, (right) nine-spot model. The parameters of the maser distributions are listed
in Table \ref{centroidstab} and shown in the top panels (the areas of the spots are proportional to the flux density).  The middle panels show the simulated p-v diagrams. In the middle right panel, the position-velocity diagram from the simulation is shown with the data (also shown in Figure \ref{3masskepler}). In the bottom panels the simulated spectrum is plotted with the observed spectrum for each model. The nine component model has 35 parameters. These parameters were not determined by least-mean-square fitting, but chosen to show an approximate fit and to demonstrate the characteristics of the model. 
\label{centroids}}
\end{center}
\end{figure}

A critical characteristic of the MWC349A spectrum is the broad and flat component between the two distinct Gaussian peaks.  Our simplest model consisted of three components, the two relatively narrow components at 32 and $-14~\kms$ and a very broad component at the systemic velocity of $9~\kms$ (\citet{gor03} has made a similar spectral decomposition).  The parameters of this model are listed in Table \ref{centroidstab}.  The simulated spectrum is shown in Figure \ref{centroids} with the actual spectral data superimposed.  The central component must be broad enough in velocity to simulate the flat appearance near $9~\kms$, but not so broad as to extend beyond the outer components. Even with the large width of $21~\kms$, the model spectrum is not as flat as the data in the central region (see Figure 8, bottom left). The centroids determined from the simulated images as a function of velocity are shown in Figure \ref{centroids}. What happens effectively is that each component ``controls" the centroid position within its velocity range of dominance. That is, the Gaussian rolloff of the outer components towards the center of the spectrum leaves the systemic velocity feature dominant at velocities near the systemic value. (Note that the ``pulling" of the p-v diagram back towards zero at high velocities is simply the effect of the central feature's dominance at high velocities due to its uncharacteristically large linewidth). This result shows that the approximately linear p-v diagram slope in the low velocity region that we measured on MWC349A cannot be explained by a simple two component model or a more realistic three component model described here. That is, the spectral separation of the outer components is sufficient to ensure that there are no ``gradient" artifacts.

A more realistic model is one with nine components, whose parameters are listed in Table \ref{centroidstab}. The broad central component in the three-component model is replaced by three narrower component symmetrically placed in velocity and position as shown in Figure \ref{centroids}. The purpose of these components is to create the observed gradient in the central region. In addition, the components at 32 and $-14~\kms$ have each been replaced by three components to simulate the Keplerian wings. These components follow Kepler's Law (e.g., equation 2, with $\theta = 90 \arcdeg$, i.e., midline position) for a stellar mass of 20 $\msun$ and systemic velocity of $9~\kms$.  As shown in Figure 8, this model fits the p-v diagram and spectrum rather well. A key characteristic of this model is that the three components in each of the Keplerian wings are dominated by a strong component. In the centroid analysis, the strong component "pulls" the position of each of the weaker components so as to steepen the slope of the p-v diagram. 

In the nine-component model we chose line widths of 4-5 $\kms$ which is reasonable for a thermal gas at $10^4$K, where an unsaturated maser gain of about 5, narrows the thermal linewidth by a factor of $\sqrt 5$. However, a better model might be one of many more components that are intrinsically much narrower (i.e., resulting from higher maser gain of $>100$) but add up to an approximately Gaussian profile of about $5~\kms$ width. In this case the spectral range of influence of each component will be relatively small and the centroiding would not affect the p-v diagram significantly. In this situation, the spiral arm model would be a viable interpretation of the data. 

%FIGURE 8

Instances of power law indices in p-v diagrams of under-resolved images of masers calculated from PCA analyses have been reported by others in the literature. Examples of these are the CH3OH maser emission in DR(21)N by \citet{har07}, and the SiO maser in the late type star R Aquarii by \citet{cot04} 

We note that W.J.~Welch and S.~White (private communication) have also imaged MWC349A in the same transition with the BIMA array in its highest resolution mode of 0$\farcs$1. Their image is very similar to ours, with a gradient at low velocities and ``Keplerian'' wings, although the centroiding is less accurate because of the somewhat lower signal-to-noise ratio.  Although their resolution was still insufficient to resolve the masers, we are greatly encouraged that our results are consistent. In addition, \citet{mar07} presented maser images with the Plateau-de-Bure interferometer at a recent conference, which clearly shows the linear structure of the systemic features.  They also measured features beyond our range ($>42,~<-21~\kms$) which are offset from the linear structure and may be due to a wind component. As can be seen in figure 8 (lower right panel) our spectrum shows significant power beyond $40$ and $-21~\kms$ that was too weak for us to find accurate positions for, which could be part of a wind component. 

\section{Conclusions}

We have presented a new high resolution image of the maser features in the inner ionized envelop of MWC349A.  The masers act as tracers of the velocity distribution in the disk.  We used the PCA method to determine the positions of the maser spots to an accuracy much finer than the resolution of the interferometer. Based on our simulations we find that there are two plausible interpretations to our data. In both models the linear distribution of masers in the p-v diagram between $-10$ and $25~\kms$ suggests the presence of an annular ring of maser emission with a radius of about $0\farcs025$. The scatter is due to measurement noise, intrinsic thickness in the annular masing region, and inclination of the ring.  If the high velocity maser components have linewidths of about $5~\kms$, then the simulations of the PCA method indicate that they may lie along the midline. If, on the other hand, these components have much narrower intrinsic linewidths because of higher maser amplification, then they may be distributed along spiral structures. 

If the maser components are the result of the unsaturated amplification, then their brightness temperatures will be much greater than the thermal temperature of the gas by a factor of exp(G), where G is the maser gain factor. Thus, these masers may be observable at much higher resolution. ALMA, in its extended configuration of 10 km, could be used to observe MWC349 (which will have a maximum elevation angle of about 30$\arcdeg$) at a resolution of about $0\farcs 02$ at 230 GHz, and could distinguish between the models we have proposed. 
\acknowledgments

We thank Henrik Beuther, Chunhua Qi, David Fong, Mark Gurwell, and Jun-Hui Zhao for help with the preparation of the observing script as well as the data calibration and reduction.   We are grateful to the astute remarks of the anonymous referee who urged us to look more deeply into the simulations reported here and made other very useful suggestions.  We thank William Danchi for providing digital data for Figures 1 and 4, and Yolanda G\'{o}mez and Daniel Tofoya for providing the digital image for Figure 1. Jack Welch encouraged us to carry out a simulation of centroiding analysis and showed us the BIMA images of MWC349A, and Vladimir Strelnitski discussed radiative transfer issues and dynamical models with us.  We also benefited from discussions with Naama Dror, Yolanda G\'{o}mez, Luis Rodr\'{\i}guez, Howard Smith and Daniel Tafoya. The Submillimeter Array is a joint project between the Smithsonian Astrophysical Observatory and the Academia
Sinica Institute of Astronomy and Astrophysics, and is funded by the Smithsonian Institution and the Academia Sinica.

Facilities: \facility{SMA}

%\appendix

\end{document}